\documentstyle[psbox,psfig]{WORKSHOP}
\def\arcmin{\ifmmode ^{\prime}\else$^{\prime}$\fi}
\def\arcsec{\ifmmode ^{\prime\prime}\else$^{\prime\prime}$\fi}
\def\approxlt{\mathrel{\hbox{\rlap{\lower.55ex \hbox {$\sim$}}
        \kern-.3em \raise.4ex \hbox{$<$}}}}
\def\approxgt{\mathrel{\hbox{\rlap{\lower.55ex \hbox {$\sim$}}
        \kern-.3em \raise.4ex \hbox{$>$}}}}

\begin{document}

\title{X-ray outbursts from nearby `normal' and active galaxies\\[0.2cm]
       {\Large\sf a review, new radio observations, and an X-ray search for further \\
         tidal disruption flares} }  

\author{
Stefanie Komossa$^1$ and Michael Dahlem$^2$  
\\[12pt]  
%
$^1$  Max-Planck-Institut f\"ur extraterrestrische Physik, Giessenbachstr., D-85748 Garching, Germany\\
$^2$  European Southern Observatory, Casilla 19001, Santiago 19, Chile \\
%
{\it E-mail(StK): skomossa@mpe.mpg.de} 
}

\abst{
In the last few years, giant-amplitude, non-recurrent X-ray flares have been
observed from several {\em non}-active galaxies (NGC 5905, RXJ1242-11, RXJ1624+75, RXJ1420+53,
RXJ1331-32). 
All of them share similar
properties, namely:
extreme X-ray softness in outburst, huge peak luminosity (up to $\sim 10^{44}$ erg/s),
and the absence of optical signs of Seyfert activity.
Tidal disruption of a star by a supermassive black hole is the
favored explanation of these unusual events. 

We present a review of the previous results,  
a search for radio emission from all outbursters, based on the NVSS database,
and dedicated radio observations of NGC\,5905
carried out with the VLA. These provide important constraints on 
the presence of an (obscured) active nucleus (AGN) at the center of each flaring galaxy.   

We rigorously explore AGN scenarios to account for the unusual
X-ray outbursts from the optically `normal' galaxies and find
AGN-related models highly unlikely. 
We conclude that the previously favored scenario -- tidal disruption 
of a star by a supermassive
black hole at the center of each of the outbursters -- 
provides the best
explanation for the X-ray observations. 

Finally, we present results from our on-going search  
for further X-ray flares
from a sample of $\sim$140 nearby active and non-active galaxies,
using the {\sl ROSAT} data base.
While we do not find another X-ray flaring normal galaxy among
this sample -- entirely consistent with the predictions of the
tidal disruption scenario -- several highly variable {\em active}
galaxies are detected. Their variability is not linked to
tidal disruption, but best explained in terms of absorption
or accretion-disk-related models.  
}

\kword{galaxies: X-ray flares --- tidal disruption --- non-active galaxies ---
                      active galaxies --- radio observations --- individual objects: NGC 5905,
                      RXJ1242-1119, RXJ1624+7554, RXJ1420+5334, RXJ1331-3243}

\maketitle
\thispagestyle{empty}

\vspace*{-19.1cm}
\begin{verbatim}
To appear in: `MAXI workshop on AGN variability' (held in Nikko, March 2001)
Preprint available at http://www.xray.mpe.mpg.de/~skomossa
\end{verbatim}
\vspace*{17.1cm}

\section{\bf The search for supermassive black holes at the centers of galaxies, 
and flares from tidally disrupted stars as probes}

There is strong evidence for the presence of massive dark objects
at the centers of many galaxies. Does this hold for {\em all} galaxies ?  
Questions of particular interest in the context of AGN evolution are:
what fraction of galaxies have passed through an active phase,
and how many now have non-accreting and hence unseen 
supermassive black holes (SMBHs) at their centers
(e.g., Rees 1989)?

Several approaches were followed to study these questions.
Much effort has
concentrated on deriving central object masses from studies of the {\sl dynamics of
stars and gas} in the nuclei of nearby galaxies.
Earlier (ground-based) evidence for central quiescent dark masses
in non-active galaxies
has been strengthened
by recent HST results
(see Kormendy \& Richstone 1995 for
a review).
There is now excellent evidence for a SMBH in our galactic center
as well (Eckart \& Genzel 1996).

Whereas the dynamics of stars and gas probe rather large 
volumes, i.e., distances from the SMBH, 
high-energy {\sl X-ray emission} 
originates from the very vicinity of
the SMBH (see Komossa 2001 for a review).
In {\em active} galaxies, excellent evidence for the presence of SMBHs
is provided by the detection of luminous hard power-law like X-ray emission,
rapid variability, and the detection of relativistically broadened FeK$\alpha$ lines
(e.g., Tanaka et al. 1995). 
How can we find {\em dormant} SMBHs in {\em non-active} galaxies ?
Lidskii \& Ozernoi (1979) and Rees (1988, 1990)
suggested to use the flare of electromagnetic radiation produced
when a star is tidally disrupted and accreted by a SMBH
as a means to detect SMBHs in nearby non-active galaxies.

Historically, tidal disruption of stars by black holes was first considered
in relation to star clusters (e.g., Frank \& Rees 1976), and was  
applied to the nuclei of active galaxies where it was suggested as a means
of fueling AGN (e.g., Hills 1975), or to explain UV-X-ray variability
of AGN (e.g., Kato \& Hoshi 1978).

Depending on its trajectory, a star gets tidally disrupted after passing a
certain distance to the black hole (e.g., Hills 1975, Lidskii \& Ozernoi 1979,
Diener et al. 1997),
the tidal radius, given by
\begin{equation}
r_{\rm t} \approx r_* ({M_{\rm BH}\over M_*})^{1 \over 3} ~~.
\end{equation}
The star is first heavily distorted, then disrupted.
About 75\% of the gaseous debris becomes unbound and is 
lost from the system (e.g., Young et al. 1977, Ayal et al. 2000).
The rest will eventually be accreted by the black hole
(e.g., Cannizzo et al. 1990, Loeb \& Ulmer 1997).
The debris, first spread over a number of orbits,
quickly circularizes (e.g., Rees 1988, Cannizzo et al. 1990)
due to the action of strong
shocks when the most tightly bound debris interacts with
other parts of the stream (e.g., Kim et al. 1999).
Most orbital periods will then be within a few times
the period of the most tightly bound matter
(e.g., Evans \& Kochanek 1989; see also Nolthenius \& Katz 1982, Luminet \& Marck
1985).

A star will only be disrupted if its tidal radius
lies outside the Schwarzschild radius of the black hole, else
it is swallowed as a whole (this happens for black hole masses larger than 
a few $\times$ 10$^7$ M$_{\odot}$; in case of a Kerr black hole, tidal
disruption may occur even for larger BH masses if the star
approaches from a favorable direction (Beloborodov et al. 1992)).
Larger BH masses may still strip the atmospheres of giant stars.
Most theoretical work so far focussed on stars of solar mass and 
radius.{\footnote{Numerical simulations of the disruption process,
the stream-stream collision, the accretion phase, the change in angular
momentum of the black hole, the changes in the stellar distribution
of the surroundings, and the disruption rates have been studied in the
literature (e.g., Nduka 1971, Masshoon 1975, Nolthenius \& Katz 1982, 1983,
Carter \& Luminet 1985, Luminet \& Marck 1985, Evans \& Kochanek 1989,
Laguna et al. 1993, Diener et al. 1997, Ayal et al. 2000, Ivanov \& Novikov 2001; 
Lee et al. 1995, Kim et al. 1999;
Hills et al. 1975, Gurzadyan \& Ozernoi 1979, 1980, Cannizzo et al. 1990, Loeb \& Ulmer 1997,
Ulmer et al. 1998;
Beloborodov et al. 1992; Frank \& Rees 1976, Rauch \& Ingalls 1998, Rauch 1999;
Syer \& Ulmer 1999, Magorrian \& Tremaine 1999).
DiStefano et al. (2001) recently considered the case of $M > M_{\odot}$, and suggested
that some ultra-soft X-ray sources (like the one at the center of M31)
could be the remnants of tidally stripped stars.}} 

Explicit predictions of the emitted spectrum and luminosity
during the disruption process and the start of the accretion
phase are still rare (see Sect. 4.2 for details).
The emission is likely peaked in the soft X-ray or UV portion
of the spectrum, initially (e.g., Rees 1988,
Kim et al. 1999, Cannizzo et al. 1990; see also Sembay \& West 1993).

\begin{table*}[t]
\caption{Summary of the X-ray properties of the flaring normal
galaxies during outburst (NGC\,5905: Bade et al. 1996, Komossa \& Bade 1999,
RXJ1242-1119: Komossa \& Greiner 1999, RXJ1624+7554: Grupe et al. 1999,
RXJ1420+5334: Greiner et al. 2000; for first results on another 
candidate see Reiprich \& Greiner 2001. Based on the position they report,
we refer to this source as RXJ1331-3243). 
$z$ gives the redshift,  
 $T_{\rm bb}$ is the black body
              temperature derived from a black body fit to the data
 (cold absorption was fixed to the Galactic value in the direction
 of the individual galaxies). $L_{\rm x,bb}$ gives the intrinsic luminosity in the
              (0.1--2.4) keV band, based on the black body fit. 
 (We note that this is a lower limit to the actual peak luminosity,
  since we most likely have not caught the sources exactly at maximum
light, since the spectrum may extend into the EUV, and since
we have conservatively assumed no X-ray absorption intrinsic to
  the galaxies).  The last column summarizes our results on the radio properties of the galaxies. }
\vskip0.2cm
\begin{tabular}{ccccl}
  \noalign{\smallskip}
  \hline
  \noalign{\smallskip}
galaxy name & $z$ & $kT_{\rm bb}$ [keV] & $L_{\rm x,bb}$ [erg/s] & radio results \\
  \noalign{\smallskip}
  \hline
  \hline
  \noalign{\smallskip}
NGC\,5905 & 0.011 & 0.06 & 3 10$^{42}$$^*$ & no 8.5\,GHz core source: $f<0.15\,$mJy $\rightarrow$ $L < 10^{20}$ W/Hz \\
  \noalign{\smallskip}
          &       &                 &                  &  extended NVSS emission at 1.4GHz, flux $f_{1.4}$ = 21.4 mJy      \\
  \noalign{\smallskip}
RXJ1242$-$1119 & 0.050 & 0.06 & 9 10$^{43}$~ & no NVSS detection (closest source is 4.6$\arcmin$ away) \\
  \noalign{\smallskip}
RXJ1624+7554 & 0.064 & 0.097 & $\sim$ 10$^{44}$~ & no NVSS detection (closest source is 7.1$\arcmin$ away)$^{**}$ \\
  \noalign{\smallskip}
RXJ1420+5334 & 0.147 & 0.04 & 8 10$^{43}$~ &  no NVSS detection (closest source is 3.3$\arcmin$ away)\\
  \noalign{\smallskip}
RXJ1331+3243 & 0.051 &  &  &  no NVSS detection (closest source is 89$\arcsec$ away)\\
  \noalign{\smallskip}
\hline
\end{tabular}
\vskip0.15cm
  \noindent{\scriptsize $^{*}$Mean luminosity during the outburst; since the flux
 varied by a factor $\sim$3 during the observation, the peak luminosity is 
higher. $^{**}$Absence of NVSS source
 was already reported by Grupe et al. 1999.}
\end{table*}

\section{\bf Tidal disruption flares from {\itshape non-active} galaxies}

With the X-ray satellite {\sl ROSAT} (Tr\"umper 1983), 
some rather unusual observations have been made in the last few
years: the detections of giant-amplitude, non-recurrent X-ray
outbursts from a handful of {\em optically non-active} galaxies,
starting with the case of NGC\,5905 (Bade et al. 1996, Komossa \& Bade 1999). 
Based on the huge observed outburst luminosity,
the observations were interpreted in terms of tidal disruption events. 
Below, we first give a brief review of all published
X-ray flaring non-active galaxies{\footnote{Although 
not discussed in detail here, we note that during the last several years,
tidal disruption was also occasionally invoked to explain some
peculiar properties of {\em active} galaxies, although alternative
interpretations existed in each case:
Tidal disruption was applied by Eracleous et al. (1995) in a duty cycle model
to explain the UV brightness/darkness of LINERs.
Peterson \& Ferland (1986) suggested this mechanism as possible explanation for
the transient brightening and broadening of the HeII line observed in
the Seyfert galaxy NGC\,5548.
Variability in the Balmer lines of some AGN (the appearance and disappearance
of a broad component in H$\beta$ or H$\alpha$) has recently been interpreted in the
same way.
Brandt et al. (1995) reported the detection of an X-ray outburst
from the galaxy IC\,3599 (Zwicky 159.034). Besides other outburst
mechanisms, tidal disruption was briefly mentioned as possibility
(see also Grupe et al. 1995). Based on high-resolution post-outburst optical spectra,
Komossa \& Bade (1999) classified IC3599 as Seyfert type 1.9. 
In the UV spectral region, two UV spikes were detected at and near
the center of the elliptical
galaxy NGC\,4552. The central flare was interpreted by Renzini et al. 
(1995) as accretion event (the tidal stripping
of a star's atmosphere by a SMBH, or the accretion of a molecular cloud).
There are several indications (e.g., from radio observations), that 
NGC\,4552 shows permanent low-level activity  
 (see Komossa 1999
for a more complete review on this suject).}}, 
and then present new radio observations.  

{\mbox{There are now four X-ray flaring `normal' galaxies}
{\mbox{(NGC\,5905, RXJ1242-1119, RXJ1624+7554, RXJ1420+}  5334{\footnote{The
X-ray position error circle of RXJ1420+53 contains a second galaxy for which a spectrum
is not yet available. Based on the galaxy's morphology, Greiner et al (2000) argue
that it is likely non-active}}) , and a possible fifth
candidate (RXJ1331-3243), all of which show similar 
properties: 
\begin{itemize}

\item
huge X-ray peak luminosity (up to $\sim 10^{44}$ erg/s),

\item
giant amplitude of variability (up to a factor $\sim$ 200),

\item
ultra-soft X-ray spectrum ($kT_{\rm bb} \simeq$ 0.04-0.1 keV
when a black body model is applied), 

\item
absence of optical signs of Seyfert activity (the spectrum of NGC\,5905
is of HII-type; the other galaxies do not show any emission lines). 
 
\end{itemize}

\noindent A summary of the observations is provided in Table 1. 
In Fig. 1 we have overplotted the X-ray lightcurves
of NGC\,5905 and RXJ1420+53, shifted in time to the
same date of outburst to allow direct comparison. 
So far, the best sampled lightcurve is that of NGC\,5905.
The `merged' lightcurve is consistent with a fast rise
and a decline on a time scale of months to years.

\subsection{\bf Radio observations}

A potential alternative to the favored tidal disruption scenario 
is the presence of a peculiar, optically hidden AGN at the center of
each flaring galaxy. As already discussed by Komossa \& Bade (1999)
this possibility is unlikely. It is very important, though, to {\em exclude}
the presence of an AGN. Besides hard X-ray observations,
compact radio emission is a good indicator of AGN activity 
because radio photons can penetrate even high-column density
dusty gas which is not transparent to optical or soft X-ray photons.

\subsubsection{NVSS search for radio emission from the X-ray outbursters}

We have performed a search for radio emission from the X-ray flaring
galaxies. 
We used the NRAO VLA Sky Survey (NVSS) catalogue (Condon et al. 1998) which contains
the results of a 1.4\,GHz radio sky survey north of $\delta$=--40$^{\rm o}$. 
The survey reaches a limiting source brightness of $\sim$2.5mJy/beam.   
Except NGC\,5905, no flaring galaxy has a NVSS detection.
The emission of NGC\,5905 appears extended and is thus likely related
to the galaxy instead of the nucleus (see also next Section).  

\subsubsection{VLA 8.5\,GHz observations of NGC\,5905}

In order to search for a radio source at the nucleus of NGC\,5905
we have carried out a radio observation with the VLA A array at 8.46 GHz.
The observation was performed on November 3, 
1996 with a duration
of 2380 sec. The band width was 100 MHz. A resolution of about 0.2$\arcsec$
was achieved.   

No radio source is detected within the central field 
of view of 100$\arcsec$$\times$100$\arcsec$. 
Based on the background noise level, we derive a 5$\sigma$ upper limit
for the presence of a central point source of 0.15 mJy. 
Assuming a distance of 75.4 Mpc of NGC\,5905 this
translates into an upper limit on the luminosity
of $L_{\rm 8.46GHz} \leq\ 1.0\,10^{20}$ W/Hz.

\vskip0.4cm

\section{\bf Outburst scenarios}

\subsection{\bf Alternatives to tidal disruption}


\subsubsection{Stellar sources, lensing, GRBs}

Firstly, we note that based purely on a positional coincidence,
interlopers (flaring Galactic foreground objects)
could not be completely excluded, given the
limited spatial positional accuracy of {\sl ROSAT} of at
least several arcseconds. 
However, known populations of galactic flaring sources
show different temporal properties.
Furthermore, in the case of the nearby galaxy
NGC\,5905 we can clearly locate the X-ray emission at
the nuclear region of this galaxy.{\footnote{ 
In the other cases, X-ray error circles are larger.
The superb spatial resolution of {\sl Chandra} will provide a crucial
test, by precisely locating the post-flare X-ray emission
which should coincide with the nucleus of each flaring galaxy.}

Other sources of the X-ray emission
related to sources within the galaxies
NGC\,5905 and RXJ1242--11
were reviewed by Komossa \& Bade (1999)
in some detail, including some order of
magnitude estimates:
Most outburst scenarios do not survive
close scrutiny, because they
cannot account for the huge maximum luminosity (e.g.,
X-ray binaries within the galaxies, or a supernova in a dense medium),
are inconsistent
with the optical observations (gravitational lensing),
or predict
a different temporal behavior (X-ray afterglow of a Gamma-ray burst).

\begin{figure}[t]
\psbox[xsize=8.5cm]
{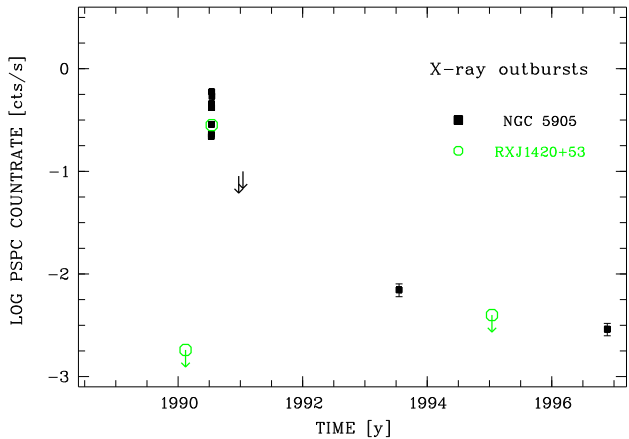}
\caption{X-ray light curve of NGC\,5905 (black squares), and RXJ1420+53
(open circles; shifted in time to match outburst date of NGC\,5905). Arrows
denote upper limits.}
\end{figure}

\subsubsection{AGN-related scenarios}

Standard AGN scenarios cannot
account for the X-ray flares and the absence of optical
AGN-like emission lines (Komossa \& Bade 1999).
Below, we describe more complicated AGN scenarios, 
and why we consider them unlikely. 

{\paragraph{\mbox{Scenario (1): pure source varia\-bility;}} accre\-tion-disk in\-stability in an LLAGN.}}
Basic idea: We have a {\em direct view} on the central
engine. The observed variability
is caused by some accretion-disk instability mechanism (e.g., Honma et al. 1991).
In that case, we already know: if there is an AGN at all in NGC 5905,
it is a  {\sl low-luminosity} AGN (LLAGN)
because $L_{\rm x,low-state} \approx 4\,10^{40}$ erg/s is observed.
Such luminosities have been seen in LINERs, but they are not
strongly X-ray variable (e.g., Komossa et al. 1999).
A similar number ($10^{40-41}$ erg/s) should hold for 
RXJ1242-11 and the other outbursters because,  
if they had 10$^{42}$ erg/s or more in low-state, we should have detected
NLR emission lines.
A low-state luminosity of $10^{40-41}$ erg/s would make
the total amplitude of variability of RXJ1242-11,RXJ1624+75, RXJ1420+53, and
RXJ1331-32 giant: a factor $10^{3-4}$.
When trying to explain this with AGN-related scenarios, it has to be kept in mind
that such variability has never been observed in known AGN.

\paragraph{Scenario (2): pure absorption variability; 
                orbiting cold, dusty absorber with a `hole'.}
Basic scenario: The AGN is absorbed by a cold absorber in nearly all directions.
If this absorber has a hole, and if this hole passes
our line-of-sight,
we have a short
view on the intrinsic AGN, thus see a flare.
Instead of an `orbiting hole', an ensemble of clouds, covering
the intrinsic source most of the time, might produce
occasional non-shadowing of the central source.
Such a scenario is invoked by Risaliti \& Elvis (2001) to
explain the variable cold absorption detected in many Seyfert\,2 galaxies.
In order to shield the NLR totally,
the cold material would have to be dusty and cover nearly 4$\pi$.
However, this model does not explain the extreme X-ray softness of 
the outbursters. If we have a short glimpse on a `normal'
AGN, we would expect to see a more typical AGN spectrum in high-state.  

\paragraph{Scenario (3): source + absorption variability;
 intrinsic source variability plus
 related variability in the ionization state of an absorber.}
 Basic idea: Presence of an AGN which is surrounded by an absorber.
 The AGN is intrinsically
 variable. In high-state, the originally cold absorber
 becomes a {\em warm} absorber. Ionized absorption then automatically
 explains the very soft
 X-ray spectrum in the {\sl ROSAT} band (see Komossa \& Bade 1999 for
 explicit spectral fits). In source low-state, the absorber is {\em cold}
 and absorption is complete in the {\sl ROSAT} band.
 Medium-amplitude source variability would then cause
high-amplitude observed variability.

This model cannot account for the absence of optical emission lines.
They can only be shielded if {\em dust} is mixed with the ionized absorber.
The model of a dusty warm absorber does no longer provide a
successful X-ray spectral fit to NGC\,5905, though (Komossa \& Bade 1999).

\subsection{\bf Tidal disruption model}

Except for GRB-related emission mechanisms, the huge peak outburst
luminosity nearly inevitably calls for the presence of a SMBH. 
This, in combination with the complete absence of any signs of AGN activity
from all wavebands, 
makes tidal disruption of a star by a SMBH
the most plausible outburst mechanism. 

Intense electromagnetic radiation will be emitted
in three phases of the disruption and accretion process:
First, during the stream-stream collision when different parts
of the bound stellar debris first interact with themselves (Rees 1988).
Kim et al. (1999) have carried out numerical simulations of this
process and find that the initial burst due to
the collision may reach a luminosity of 10$^{41}$ erg/s, under the assumption
of a BH mass of 10$^6$ M$_{\odot}$ and a star of solar mass and radius.
Secondly, radiation is emitted during the accretion of the stellar
gaseous debris. Finally, the unbound stellar debris leaving the system
may shock the surrounding interstellar matter 
and cause intense emission.

 \begin{figure}[t]
\psbox[xsize=8.8cm,clip=]
{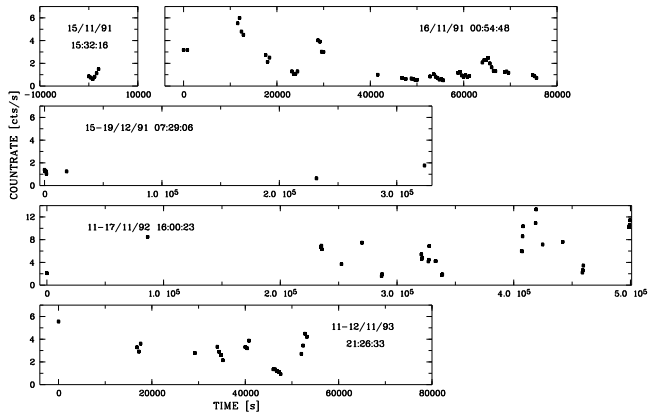}
 \caption[light]{Long-term X-ray lightcurve of NGC\,4051, based on all
pointed {\sl ROSAT} PSPC observations of this source.  NGC\,4051
is variable by a factor $\sim$30 in countrate.
The time is measured in sec from the beginning of each
observation; the inset in each panel gives the starting time. }
\label{light_4051}
\end{figure}

The luminosity emitted if the black hole is accreting at its Eddington luminosity
can be estimated by $ L_{\rm edd} \simeq 1.3 \times 10^{38} M/M_{\odot}$ erg/s.
In case of NGC 5905, a BH mass of at least $\sim 10^{5}$ M$_{\odot}$ would be
required to
produce the observed $L_{\rm x}$, and a higher mass if $L_{\rm x}$ was not observed
at its peak value.
For comparison, BH masses of $M_{\rm BH} \approxlt 10^{6 - 7} {\rm M}_\odot$
have recently been reported by Salucci et al. (2000) for the
centers of several late-type spiral galaxies.
Alternatively, the atmosphere of a giant star could have been
stripped instead of a complete disruption event.
It is interesting to note that NGC\,5905
possesses a complex bar structure (Wozniak et al. 1995, Friedli et al. 1996) which might
aid in the fueling process by disturbing the
stellar velocity fields.

Using the black body fit to the X-ray spectra
of NGC\,5905 and RXJ1242--11, we find the fiducial black
body radius
to be located between the
last stable orbit of a Schwarzschild black hole, and
inside the tidal radius.

We note that many details of the tidal disruption and the
related processes are still unclear.
In particular, the flares cannot be standardised. Observations
would depend on many parameters, like the type of disrupted star, the impact
parameter, the spin of the black hole, effects of relativistic precession,
 and the radiative transfer is complicated
by effects of viscosity and shocks (Rees 1990).
Uncertainties also include the amount of the stellar debris
that is accreted (part may be ejected as a thick wind, or
swallowed immediately). Related to this is the duration
of the flare-like activity, which may be months or years
to tens of years (e.g., Rees 1988, Cannizzo et al. 1990,
Gurzadyan \& Ozernoi 1979).

\section{\bf Search for further X-ray flares}

We performed a search for further cases of strong X-ray variability
using the sample of nearby galaxies of Ho et al. (1995) and
{\sl ROSAT} all-sky survey (Voges et al. 1999)
and archived pointed observations.
The sample of Ho et al. has the advantage of the availability
of high-quality optical spectra, which are necessary when searching
for `truly' non-active galaxies.
136 out of the 486 galaxies in the catalogue were detected
in pointed observations. For these, we compared the countrates
with those measured during the RASS.

\subsection{Non-active galaxies} 

We do not find another flaring normal galaxy.

The absence of any further flaring event among the
sample galaxies is entirely consistent with the expected
tidal disruption rate of one event in at least $\sim$10$^4$ years per galaxy
(e.g., Magorrian \& Tremaine 1999).

\subsection{AGN}

Several of the sample galaxies show variability by a factor 10--30. All
of these are well-known AGN.

Many active galactic nuclei are variable in X-rays
with a range of amplitudes, typically a factor 2--3, and on many different time scales
(e.g., Mushotzky et al. 1993).
The cause of variability is usually linked in 
one way or another to the central engine; for instance by changes in
the accretion disk (e.g., Piro et al. 1988, 1997), or by variable obscuration
(e.g., Komossa \& Fink 1997, Komossa \& Meerschweinchen 2000).

As an example for a highly variable AGN among the present 
sample galaxies, we show in Fig. 2 
the long-term {\sl ROSAT} X-ray lightcurve of NGC\,4051 
which exhibits variability in countrate of a factor $\sim$30. 
Only a small part of the variability of NGC\,4051 can be explained
with a variable warm absorber,  the rest is likely intrinsic.

Even higher total amplitude of variability is detected 
in two subsequent {\sl ROSAT} observations of NGC\,3516. The X-ray
countrate varies by a factor $\sim$50 ( Komossa \& Bade 1999,
Komossa \& Halpern 2001, in prep.); variable cold absorption
likely plays a major part in explaining the observations.

\section{\bf Future perspectives}

X-ray outbursts from non-active galaxies provide important information
on the presence of SMBHs in these galaxies,
and the link
between active and normal galaxies.
Future X-ray surveys,
like those planned with the {\sl LOBSTER} ISS X-ray all-sky monitor (Fraser 2001), {\sl MAXI} (Mihara 2001), 
and {\sl ROSITA} (Predehl 2001), 
will be valuable in finding more of these outstanding
sources.

In particular, rapid follow-up optical observations will
be important in order to detect potential emission lines
that were excited by the outburst emission.
In case a giant-amplitude X-ray flare occurs in an {\em active}
galaxy, this would also provide an excellent chance
to map the properties of the broad line region.

\vskip0.5cm
\noindent {\sl Acknowledgements:}
It is a pleasure to thank
Jules Halpern, Martin Elvis, and David L. Meier
for fruitful discussions.
The {\sl ROSAT} project has been supported by the German Bundes\-mini\-ste\-rium
f\"ur Bildung, Wissenschaft, Forschung und Technologie
(BMBF/DLR) and the Max-Planck-Society.
\\
Preprints of this and related papers can be retrieved at \\ 
http://www.xray.mpe.mpg.de/$\sim$skomossa/

\section*{References}

\re

\re Ayal S., Livio M., Piran T., 2000, ApJ 545, 772 

\re Bade N., Komossa S., Dahlem M., 1996, A\&A 309, L35

\re Beloborodov A.M., Illarionov A.F., Ivanov P.B., Polnarev A.G., 1992, MNRAS 259, 209

\re Brandt W.N., Pounds K.A., Fink H.H., 1995, MNRAS 273, L47

\re Cannizzo J.K., Lee H.M., Goodman J., 1990, ApJ 351, 38

\re Carter B., Luminet J.P., 1985, MNRAS 212, 23

\re Condon J.J., Cotton W.D., Greissen E.W., et al., 1998, AJ 115, 1693

\re Diener P., Frolov V.P., Khokhlov A.M., Novikov I.D.,
             Pethick C.J., 1997, ApJ 479, 164

\re Di\,Stefano R., Greiner R., Murray S., Garcia M., 2001, ApJL, in press

\re Eckart A., Genzel R., 1996, Nature 383, 415

\re Eracleous M., Livio M., Binette L., 1995, ApJ 445, L1

\re Evans C.R., Kochanek C.S., 1989, ApJ 346, L13

\re Frank J., Rees M.J., 1976, MNRAS 176, 633

\re Fraser G., 2001, in: MAXI workshop on AGN variability, these proceedings

\re Friedli D., Wozniak  H., Rieke M., Martinet L.,
            Bratschi P., 1996, A\&AS 118, 461

\re Greiner J., Schwarz R., Zharikov S., Orio M., 2000, A\&A 362, L25 

\re Grupe D., Beuermann K., Mannheim K., et al.,
            1995, A\&A 299, L5

\re Grupe D., Leighly K., Thomas H., 1999, A\&A 351, L30

\re Gurzadyan V.G., Ozernoi L.M., 1979, Nature 280, 214

\re Gurzadyan V.G., Ozernoi L.M., 1980, A\&A 86, 315

\re Hills J.G., 1975, Nature 254, 295

\re Ho L.C., Filippenko A.V., Sargent W.L.W., 1995, ApJS 98, 477

\re  Honma F., Matsumoto R., Kato S., 1991, PASJ 43, 147

\re Ivanov P.B., Novikov I.D., 2001, ApJ 549, 467 

\re Kato M., Hoshi R., 1978, Prog. Theor. Phys. 60/6, 1692

\re Kim S.S., Park M.-G., Lee H.M., 1999, ApJ 519, 647 

\re Komossa S., Fink H., 1997, A\&A 327, 555


\re Komossa S., Bade N., 1999, A\&A 343, 775

\re Komossa S., B\"ohringer H., Huchra J., 1999, A\&A 349, 88

\re Komossa S., Greiner J., 1999, A\&A 349, L45


\re Komossa S., 1999, in Proc: ASCA/ROSAT Workshop on~AGN and the X-ray Background,
      T. Takahashi, H. Inoue (eds), ISAS Report, p. 149;
      [also available at astro-ph/0001263]

\re Komossa S., Meerschweinchen J., 2000, A\&A 354, 411

\re Komossa S., 2001, in Proc: IX. Marcel Grossmann Meeting
 on General Relativity, Gravitation and Relativistic Field Theories,
 V. Gurzadyan et al. (eds), in press [astro-ph/0101289]

\re Kormendy J., Richstone D.O., 1995, ARA\&A 33, 581

\re Laguna P., Miller W.A., Zurek W.H., Davies M.B., 1993, ApJ 410,
              L83

\re Lee H.M., Kang H., Ryu D., 1995, ApJ 464, 131

\re Lidskii V.V., Ozernoi L.M., 1979, Sov. Astron. Lett. 5(1), 16

\re Loeb A., Ulmer A., 1997, ApJ 489, 573

\re Luminet J.P., Marck J.-A., 1985, MNRAS 212, 57

\re Magorrian J., Tremaine S., 1999, MNRAS 309, 447

\re Mashoon B., 1975, ApJ 197, 705

\re Mihara T., 2001, in: MAXI workshop on AGN variability, these proceedings

\re Mushotzky R.F., Done C., Pounds K.A., 1993, ARA\&A 31, 717

\re Nduka A., 1971, ApJ 170, 131

\re Nolthenius R.A., Katz J.I, 1982, ApJ 263, 377

\re Nolthenius R.A., Katz J.I, 1983, ApJ 269, 297

\re Peterson B.M., Ferland G.J., 1986, Nature 324, 345

\re Piro L., Massaro E., Perola G.C., Molteni D., 1988, ApJ 325, L25

\re Piro L.,  
                et al., 1997, A\&A 319, 74

\re Predehl P., 2001, in: MAXI workshop on AGN variability, these proceedings    

\re Rauch K.P., 1999, ApJ 514, 725

\re Rauch K.P., Ingalls B., 1998, MNRAS 299, 1231

\re Rees M.J., 1988, Nat 333, 523

\re Rees M.J., 1989, Rev. mod. Astr. 2, 1

\re Rees M.J., 1990, Science 247, 817

\re Reiprich T., Greiner J., 2001, in Proc: 
          ESO workshop on black holes in binaries and AGN, p. 168 

\re Renzini A., et al., 
           1995, Nature 378, 39

\re Risaliti G., Elvis M., 2001, ApJ, submitted

\re Salucci P., Ratnam C., Monaco P., Danese L., 2000, MNRAS 317, 488

\re Sembay S., West R.G., 1993, MNRAS 262, 141

\re Syer D., Ulmer A., 1999, MNRAS 306, 35

\re Tanaka  Y., et al., 1995, Nature 375, 659

\re Tr\"umper J., 1983, Adv. Space Res. 2, 241

\re Ulmer A., Paczynski B., Goodman J., 1998, A\&A 333, 379

\re Voges W., et al., 1999, A\&A 349, 389 

\re Wozniak H., Friedli D., Martinet L., Martin P., Bratschi P., 1995, ApJS 111, 115

\re Young P.,
                     Shields G.,
           Wheeler J.C., 1977, ApJ 212, 367

\label{last}

\end{document}